\documentclass[conference]{IEEEtran}
\usepackage{algorithm,algorithmicx}
\makeatletter

\usepackage{cite}
\usepackage{amsmath,amssymb,amsfonts,bm}
\usepackage{algpseudocode,fancyhdr,amsfonts,cases,graphicx,xcolor}
\usepackage{graphicx}
\usepackage{textcomp}
\usepackage{comment,flushend}
\usepackage{enumitem}
\usepackage{xcolor,comment}
\def\BibTeX{{\rm B\kern-.05em{\sc i\kern-.025em b}\kern-.08em
    T\kern-.1667em\lower.7ex\hbox{E}\kern-.125emX}}
    
\newcommand{\x}{{\mathsf{x}}}
\newcommand{\y}{{\mathsf{y}}}
\newcommand{\z}{{\mathsf{z}}}
\newcommand{\e}{\mathsf{e}}
\newcommand{\jj}{\jmath}

\def\bomega{\boldsymbol{\omega}}

\def\bp{\mathbf{p}}
\def\bv{\mathbf{v}}
\def\bu{\mathbf{u}}

\def\sW{\mathcal{W}}

\def\sP{\mathcal{P}}
\def\sM{\mathcal{M}}

\begin{document}
%\title{Codebook-Based Channel Estimation for RIS-Enabled MIMO Communication Systems}
\title{Near-Field Hierarchical Beam Management for RIS-Enabled Millimeter Wave Multi-Antenna Systems}

\author{\IEEEauthorblockN{George C. Alexandropoulos\IEEEauthorrefmark{4}, Vahid Jamali\IEEEauthorrefmark{2}, Robert Schober\IEEEauthorrefmark{1}, and H. Vincent Poor\IEEEauthorrefmark{2}}
\IEEEauthorblockA{\IEEEauthorrefmark{4}Department of Informatics and Telecommunications, National and Kapodistrian University of Athens, Greece\\
\IEEEauthorrefmark{2}Department of Electrical and Computer Engineering, Princeton University, USA\\
\IEEEauthorrefmark{1}Institute for Digital Communications, Friedrich-Alexander University Erlangen-N\"urnberg (FAU), Germany\\
Emails: alexandg@di.uoa.gr, \{jamali, poor\}@princeton.edu, robert.schober@fau.de
}\vspace{-0.8cm}}

\maketitle
\begin{abstract}
In this paper, we present a low overhead beam management approach for near-field millimeter-wave multi-antenna communication systems enabled by Reconfigurable Intelligent Surfaces (RISs). We devise a novel variable-width hierarchical phase-shift codebook suitable for both the near- and far-field of the RIS, and present a fast alignment algorithm for the RIS phase shifts and the transceiver beamformers.
%, which can be followed by conventional pilot-assisted estimation of the RIS-parametrized end-to-end channel matrix. 
Indicative performance evaluation results are shown, verifying the effectiveness of the proposed approach in comparison with various benchmark schemes. 
\end{abstract}

\begin{IEEEkeywords}
Reconfigurable intelligent surfaces, beam management, hierarchical codebook, near-field regime, millimeter wave.
\end{IEEEkeywords}

\vspace{-0.1cm}
\section{Introduction}
\vspace{-0.1cm}
The concept of programmable radio propagation environments \cite{liaskos2018new,di2019smart,wu2019towards,RISE6G_2021}, and its potential for cost- and energy-efficient smart wireless networking, is lately gaining increased interest thanks to the technology of Reconfigurable Intelligent Surfaces (RISs) \cite{huang2019reconfigurable,WavePropTCCN,risTUTORIAL2020}. Due to their minimal hardware footprint, RISs are envisioned to coat objects in the wireless medium, offering new degrees of freedom for diverse communication, localization, and sensing improvements. However, their performance gain mostly relies on the availability of Channel State Information (CSI), which given the typically large number of reflecting elements (denoted, henceforth, by $Q$) translates into a huge, and often unaffordable, CSI acquisition overhead \cite{wei2021channel,yu2021smart}. 

There exists already a remarkable amount of literature on Channel Estimation (CE) techniques for RIS-assisted wireless communications systems; see \cite{CE_overview_2022} for a recent overview. %For example, the ON/OFF protocol proposed in \cite{mishra2019channel} comprises $Q$ stages, where in each stage, only one reflecting element is activated and the corresponding cascaded Base Station (BS)-RIS-Mobile User (MU) channel is estimated. 
The two main CE categories, whose overhead does not scale with $Q$, exploit sparsity- and codebook-based schemes \cite{wang2020compressed,jamali2021power}. In particular, the sparsity of the wireless channel in the angular domain was exploited in \cite{wang2020compressed} to design a CE algorithm, whose overhead scales with the number of dominant propagation paths of the wireless channel. In contrast, in \cite{jamali2021power}, the authors proposed to estimate the end-to-end Base-Station (BS) to RIS to Mobile User (MU) channel only for a limited number of RIS phase-shift configurations drawn from a codebook. The presented small-sized phase-shift design led to CE overhead scaling with the codebook size. 
%Since the CE overhead scaled with the codebook size, an RIS phase-shift profile was developed, which allowed the design of small-sized phase-shift codebooks. 
In \cite{Fast_Beam_Rui_2020}, a multi-beam training method using the codebook approach of \cite{Hierarchical_mmWave_2016} was proposed, according to which the RIS elements were divided into sub-surfaces, with each realizing a distinct codebook intended for a distinct MU. A similar approach was considered in \cite{Singh_2021}, where the authors focused on minimizing beam request collisions for MUs belonging in similar impinging angles. In \cite{Ning_THz_2021}, a hierarchical beam searching scheme for RIS-assisted THz communications was presented, which relied on RIS phase-shift codebooks for efficient ternary-tree search. Random-beamforming-based maximum likelihood estimation of the parameters associated with the Line-Of-Sight (LOS) component, while treating other non-LOS components as interference, was proposed in \cite{Wang_2021}. In \cite{Hu_multicast_2020}, an uplink beam training scheme was designed for multi-antenna multicast systems, where the RIS was used as a component of the transmitter. By exploiting the inherent sparse structure of BS-RIS-MU cascade millimeter-Wave (mmWave) channels, in \cite{Wang_THz_2021}, multi-directional beam training sequences were presented, while the beam training problem was formulated as a joint sparse sensing and phaseless estimation problem. In \cite{cai_Hierarchical_2021}, a hierarchical RIS phase-shift configuration design was presented, together with an algorithm that exploits the angular-domain channel representation, to maximize the MU achievable rate. CE with hierarchical beam searching was also considered in \cite{zegrar_General_2020}, focusing on the channel between the RIS and MU.

All aforementioned works, similar to most available codebook-based CE schemes for RIS-assisted wireless systems, consider signal propagation in the far-field of the RIS. However, for high-frequency scenarios incorporating large RISs for confronting high pathloss attenuation, and thus, offering cost- and energy-efficient coverage extension, the wireless communication will often take place in the near-field of RISs. Motivated by this fact, in this paper, we present a beam management framework for RIS-enabled mmWave multi-antenna communication systems. Capitalizing on the wide illumination approach recently designed in \cite{jamali2021low} for low-overhead CE in mobile scenarios, we devise a variable-width hierarchical phase-shift codebook for near-field communications. Our simulation results for a typical mmWave communication system confirm that the proposed hierarchical beam management algorithm is able to successfully find a suitable RIS phase-shift configuration for extremely large IRSs (with $Q\approx 8000$ reflecting elements) with only a few pilot transmissions (i.e., $24$), while achieving performance close to that of a benchmark that assumes the availability of full CSI (i.e., $\approx 16,000$ channel coefficients).

\begin{figure*}[!t]
		\centering
		\includegraphics[width=1.4\columnwidth]{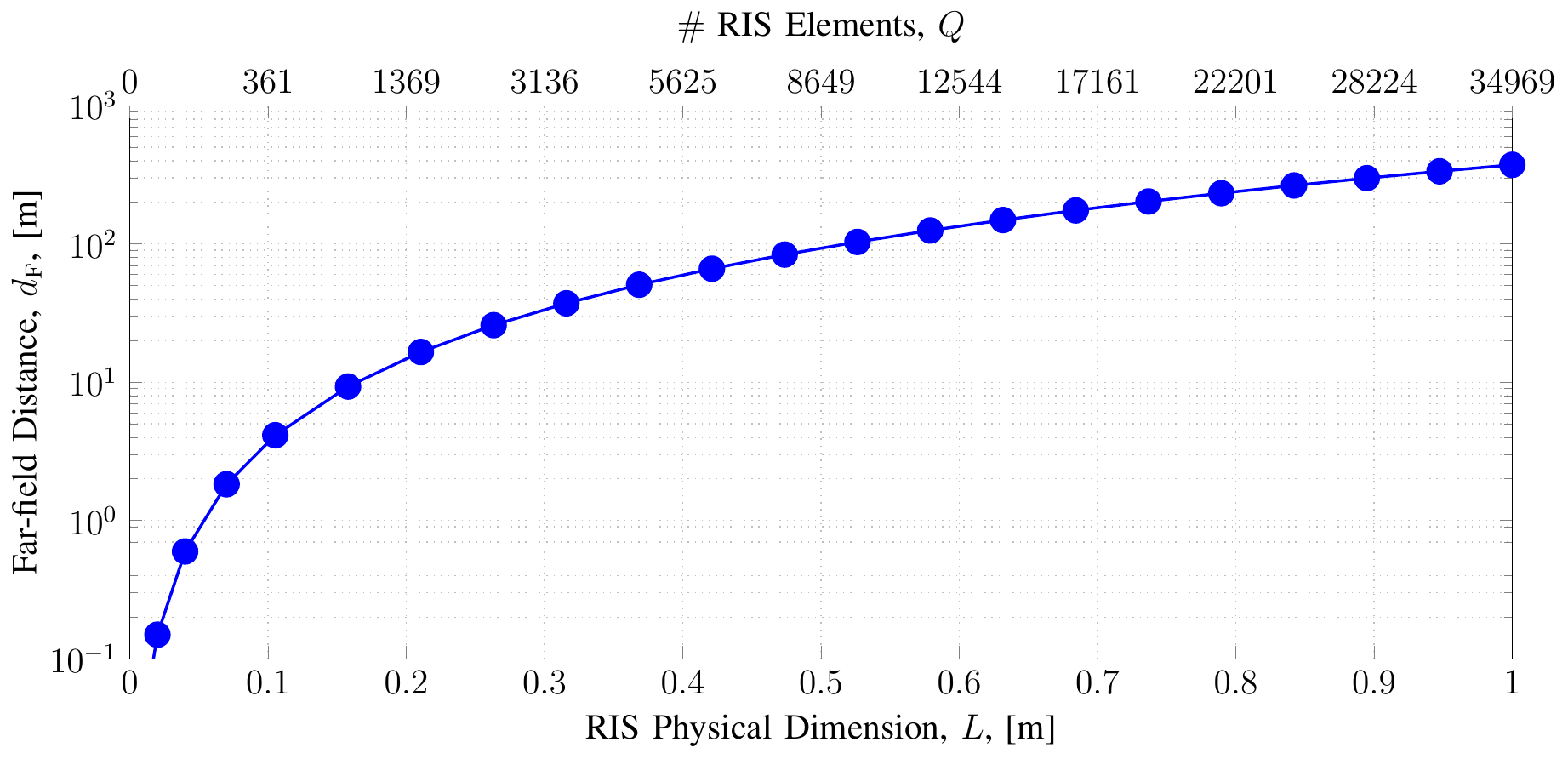}\hspace{-0.18cm}
		\caption{Far-field distance versus the RIS physical dimension $L_\y=L_\z\triangleq L$ (i.e., $D=(L_\y^2+L_\z^2)^{1/2}=\sqrt{2}L$) and the number of its unit elements $Q$ for $28$ GHz carrier frequency and element spacing $d_\y=d_\z\triangleq d=\lambda/2$.}\label{fig:farfield}	
\end{figure*} 

\subsubsection*{Notation} Vectors and matrices are denoted by boldface lowercase and boldface capital letters, respectively. $\mathbf{A}^{\rm H}$ is $\mathbf{A}$'s Hermitian transpose, $\mathbf{I}_{n}$ ($n\geq2$) is the $n\times n$ identity matrix, and $\mathbf{0}_{n\times m}$ ($n,m\geq1$) is an $n\times m$ matrix of zeros. $\left\|\mathbf{a}\right\|$ denotes $\mathbf{a}$'s Euclidean norm, $\mathbb{C}$ represents the complex number set, $\mathbb{E}\{\cdot\}$ is the expectation operator, and $\mathbf{x}\sim\mathcal{CN}(\mathbf{a},\mathbf{A})$ indicates a complex Gaussian random vector with mean $\mathbf{a}$ and covariance matrix $\mathbf{A}$. Finally, $\jmath\triangleq\sqrt{-1}$ is the imaginary unit.

\vspace{-0.1cm}
\section{System and Channel Models}
\vspace{-0.1cm}
\begin{comment}
\begin{figure}[t]
	\centering
	\includegraphics[width=0.8\columnwidth]{Fig/SystemModel.pdf}
	\caption{Schematic illustration of the considered RIS-assisted downlink communication system. Throughout the paper, we mainly use the coordinate system $\x\y\z$ and employ the coordinate system $\bar{\x}\bar{\y}\bar{\z}$ to only represent the positions of the multi-antenna BS, $Q$-element RIS, and single-antenna MU. %The generic notations $\mathbf{p}_i$ and $\bar{\mathbf{p}}_i$ refer to points in the $\x\y\z$ and $\bar{\x}\bar{\y}\bar{\z}$ systems, respectively.
	}\vspace{-0.5cm}\label{fig:system_model}
\end{figure}
\end{comment}
\subsection{System Model}
\vspace{-0.1cm}
We consider a communication system model comprising one BS equipped with $N_{\rm bs}$ antenna elements, one RIS consisting of $Q\triangleq Q_{\y}Q_\z$ phase-tunable unit cell elements with inter-element spacing $d_\y$ and $d_\z$ on the $\y$- and $\z$-axis, respectively, indexed by $q_\y=0,1,\ldots,Q_\y-1$ and $q_\z=0,1,\ldots,Q_\z-1$, and one MU with $N_{\rm mu}$ antennas. In the downlink communication direction, the baseband received signal, $\mathbf{y}_{\rm mu}\in \mathbb{C}^{N_{\rm mu}\times 1}$, at the MU antenna elements can be mathematically expressed as follows:
\begin{IEEEeqnarray}{ll}\label{eq:System_Model_DL}
\mathbf{y}_{\rm mu} = \left(\mathbf{H}+\mathbf{H}_2\mathbf{\Omega}\mathbf{H}_1\right)\mathbf{x}_{\rm bs}+\mathbf{n}_{\rm mu},
\end{IEEEeqnarray}
where $\mathbf{x}_{\rm bs}\in \mathbb{C}^{N_{\rm bs}\times 1}$ denotes the transmitted signal from the BS, such that $\mathbb{E}\{\|\mathbf{x}_{\rm bs}\|^2\}\leq P_{\rm bs}$ with $P_{\rm bs}$ being the total transmit power budget of the BS, and $\mathbf{n}_{\rm mu}\sim\mathcal{CN}(\mathbf{0},\sigma^{2}_{\rm mu}\mathbf{I}_{N_{\rm mu}})$ represents the Additive White Gaussian Noise (AWGN) at the MU receiver. We assume linear precoding, i.e., $\mathbf{x}_{\rm bs}\triangleq\mathbf{v}_{\rm bs}s$, where $\mathbf{v}_{\rm bs}\in \mathbb{C}^{N_{\rm bs}}$ is the BS precoder and $s$ denotes the complex-valued data or pilot symbol; in the former case, its chosen from a discrete modulation set. The $Q\times Q$ diagonal matrix $\mathbf{\Omega}$ in \eqref{eq:System_Model_DL} is defined as $\mathbf{\Omega}\triangleq{\rm diag}\left([g\e^{\jj\omega_1}\,g\e^{\jj\omega_2}\,\ldots\,g\e^{\jj\omega_Q}]\right)$, where each $\omega_q$ (with $q=1,2,\ldots,Q$) is the phase shift applied by the $q$-th RIS unit cell element and $g\triangleq4\pi d_{\y}d_{\z}\lambda^{-2}$ is a unit-less factor resulting from each element with $\lambda$ being the wavelength \cite{jamali2021power}. Furthermore, $\mathbf{H}\in \mathbb{C}^{N_{\rm mu}\times N_{\rm bs}}$, $\mathbf{H}_1\in \mathbb{C}^{Q\times N_{\rm bs}}$, and $\mathbf{H}_2\in \mathbb{C}^{N_{\rm mu}\times Q}$ denote the channel matrices for the BS-MU, BS-RIS, and RIS-MU links, respectively. We also assume that the MU is capable of realizing fully digital combining of the received signal $\mathbf{y}_{\rm mu}$. This combining operation is accomplished by the vector $\mathbf{u}_{\rm mu}\in \mathbb{C}^{N_{\rm mu}}$ and we assume that $\|\mathbf{u}_{\rm mu}\|=1$. 

%In this paper, we assume time division duplexing and reciprocal wireless channels, and consider both the downlink and uplink directions for deciding the beamformers for the BS and MU as well as the RIS phase configuration, via the transmission of pilot signals. Let  $\mathbf{v}_{\rm mu}\in \mathbb{C}^{N_{\rm mu}\times 1}$ denote the precoding vector for the pilot symbol $q$ sent in the uplink by the MU. We assume that $\mathbb{E}\{\|\mathbf{v}_{\rm mu}q\|^2\}\leq P_{\rm mu}$, where $P_{\rm mu}$ is the total MU transmit power. Similar to \eqref{eq:System_Model_DL}, the received pilot signal, $\mathbf{y}_{\rm bs}\in \mathbb{C}^{N_{\rm bs}\times 1}$, at the BS antennas can be expressed as
%\begin{IEEEeqnarray}{ll}\label{eq:System_Model_UL}
%\mathbf{y}_{\rm bs} = \left(\mathbf{H}^{\rm H}+\mathbf{H}_1^{\rm H}\mathbf{\Omega}\mathbf{H}_2^{\rm H}\right)\mathbf{v}_{\rm mu}q+\mathbf{n}_{\rm bs},
%\end{IEEEeqnarray}
%where $\mathbf{n}_{\rm bs}\in \mathbb{C}^{N_{\rm bs}\times 1}$ represents the AWGN at the BS receiver, which is assumed distributed as $\mathcal{CN}(\mathbf{0},\sigma^{2}_{\rm bs}\mathbf{I}_{N_{\rm bs}})$. The BS then processes $\mathbf{y}_{\rm bs}$ with the digital combiner $\mathbf{u}_{\rm bs}\in \mathbb{C}^{N_{\rm bs}\times 1}$.

\subsection{Near-Field Channel Model}
\vspace{-0.1cm}
A pragmatic criterion that quantifies the transition between the far-field and near-field regimes is the critical distance, beyond which the maximum phase error caused by the assumption of the linear phase wavefront in the far-field does not exceed $\pi/8$ \cite[Ch.~4]{balanis2015antenna}. We define this distance as $d_{\rm F}\triangleq \frac{2D^2}{\lambda}=2\bar{D}\lambda$, with $D$ and $\bar{D}$ being the largest dimension of the RIS in meters and its electric size (unit-less), respectively. Figure~\ref{fig:farfield} illustrates $d_{\rm F}$ as a function of the RIS physical size $L\triangleq L_\y=L_\z$ (i.e., $D=\sqrt{2}L$) and $Q$ (i.e., $\bar{D}=\sqrt{2Q}d/\lambda$ where $d\triangleq d_\y=d_\z=\lambda/2$). We note that $d_{\rm F}$ increases with frequency for a given $L$ (or equivalently the RIS physical dimension), but decreases with frequency when $Q$ (or equivalently the RIS electric dimension) is kept fixed.

The involved wireless channels are modeled as collections of propagation paths. Assuming $\ell_1$ paths for the BS-RIS link, the channel gain between the $m$-th antenna (with $m=1,2,\ldots,N_{\rm bs}$) at the BS and the $q$-th element at the RIS, having distance ${\rm d}_{1,i,qm}$ in the $i$-th path, can be expressed as:
\begin{IEEEeqnarray}{ll}\label{eq:nearfield}
[\mathbf{H}_1]_{q,m}\triangleq\sum_{i=0}^{\ell_1-1} {\rm PL}_{1,i}\gamma_{1,i}\e^{\jj\frac{2\pi}{\lambda}{\rm d}_{1,i,qm}},
\end{IEEEeqnarray}
where ${\rm PL}_{1,i}$ and $\gamma_{1,i}$ denote the channel pathloss and small-scale fading of the BS-RIS link, respectively. Similarly, the components of the BS-MU and RIS-MU channel matrices are given by $[\mathbf{H}]_{n,m}\triangleq \sum_{i=0}^{\ell_h-1} {\rm PL}_{h,i}\gamma_{h,i}\e^{-\jj\frac{2\pi}{\lambda}{\rm d}_{h,i,nm}}$ and $[\mathbf{H}_2]_{n,q}\triangleq \sum_{i=0}^{\ell_2-1} {\rm PL}_{2,i}\gamma_{2,i}\e^{-\jj\frac{2\pi}{\lambda}{\rm d}_{2,i,nq}}$, respectively, where the parameters are defined analogously; for example, ${\rm PL}_{2,i}$ is the  pathloss of the $i$-th path in the RIS-MU channel. 
In this paper, we consider cases where the BS-RIS and RIS-MU links have dominant LOS channel components (denoted by index $i=0$) and their distances are below $d_{\rm F}$, and the direct BS-MU link is severely blocked. For the latter, we incorporate a strong loss factor in ${\rm PL}_{h,i},\,\,\forall i$.  

\begin{figure*}[t]
		\centering
		\includegraphics[width=1.4\columnwidth]{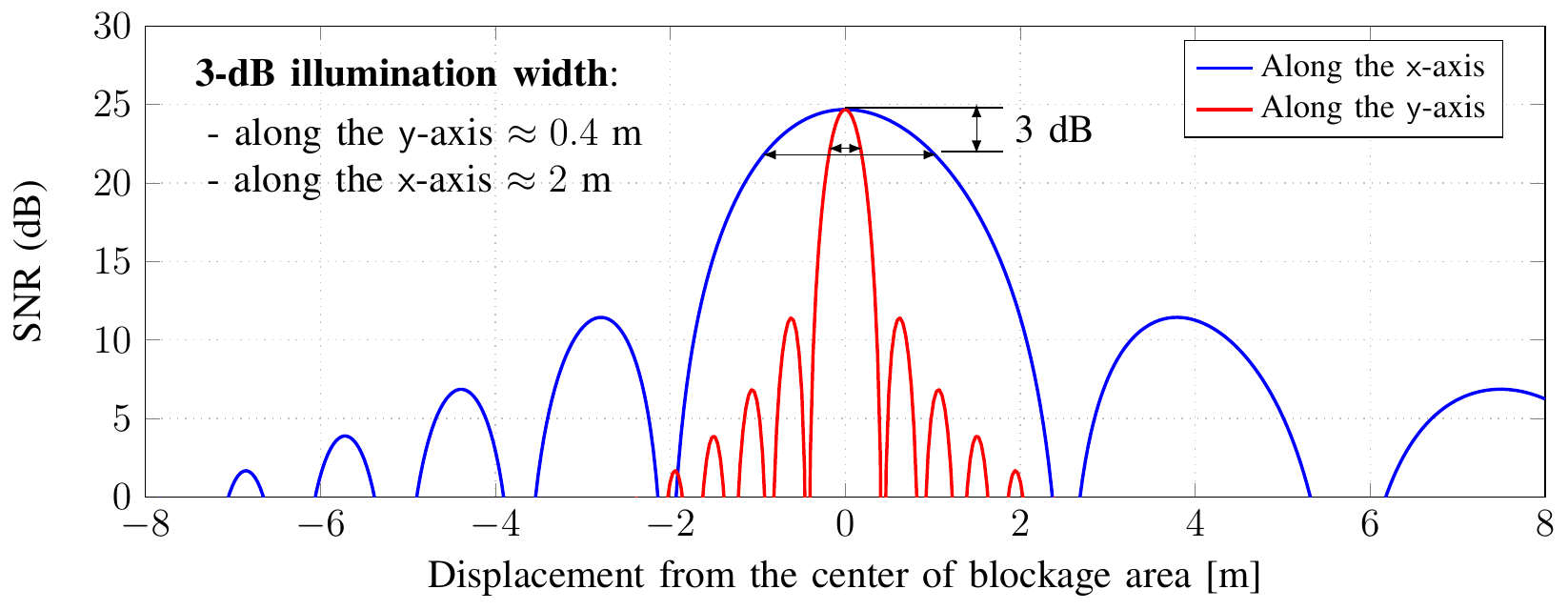}\hspace{-0.18cm}
		\caption{Received SNR vs. the displacement along the $\x$- and $\y$-axes when the RIS is configured to focus on the center of the blockage area. The system parameters are the same as those used in Section~IV. In particular, BS and RIS are located at $\bp_{i} =[40, 0, 10]$ and $\bp_{\rm ris} =[0, 40, 5]$, respectively, the focus point is at $\bp_{b} =[20, 40, 1]$,  $L_\y=L_\z=0.5$~m, $d_\y=d_\z=\frac{\lambda}{2}$, and $f=28$~GHz (i.e., $Q=8649$). }\vspace{-0.2cm}\label{fig:focus}	
\end{figure*}

%%%%%%%%%%%%%%%%%%%%%%%%%%%%%%%%%%%%%%%%%%%%%%%%%%%%%%%%%%
%\section{Hierarchical Phase Profile Design}
\vspace{-0.1cm}
\section{Proposed Beam Management Framework}
\vspace{-0.1cm}
%%%%%%%%%%%%%%%%%%%%%%%%%%%%%%%%%%%%%%%%%%%%%%%%%%%%%%%%%%
%Our objective in this section is 
The first objective of this section is to design a set of ${\rm W}$ RIS phase-shift codebooks, denoted by $\sW_1,\sW_2,\dots,\sW_{\rm W}$, where $|\sW_1|<|\sW_2|<\dots<|\sW_{\rm W}|$ and each level's codebook $\sW_w\triangleq\{\bomega_{w,1},\dots,\bomega_{w,|\sW_w|}\}$, with $w=1,2,\ldots,{\rm W}$, consists of $|\sW_w|$ RIS phase shifts $\bomega_{w,k_w}\triangleq[\omega_{w,k_w,1},\dots,\omega_{w,k_w,Q}]$ with $k_w=1,2,\ldots,|\sW_w|$. We employ the near-field phase-shift design in \cite{jamali2021low} for constructing the hierarchical codebooks in this paper. In particular, we characterize each RIS phase-shift codeword $\bomega_k$ in terms of the Generalized Radar Cross Section (GRCS) $g_{\rm ris}$ defined in \cite{najafi2020intelligent}, which is used to derive the free-space end-to-end pathloss of the RIS-enabled link, denoted by ${\rm PL}_{\rm ris}$, according to \cite[Lemma~1]{najafi2020intelligent} (we have normalized \cite{najafi2020intelligent}'s definition to $\frac{\lambda}{\sqrt{4\pi}}$ in order to get a unit-less quantity):
\begin{IEEEeqnarray}{ll}\label{eq:pathloss}
{\rm PL}_{\rm ris} \triangleq {\rm PL}_{1,0} {\rm PL}_{2,0} |g_{\rm ris}|^2,
\end{IEEEeqnarray}
where ${\rm PL}_1$ and ${\rm PL}_2$ are the free-space pathlosses of the BS-RIS and RIS-MU LOS links, respectively. The value of $g_{\rm ris}$ depends on the RIS phase-shift configuration value as well as the relative positions of the BS and MU with respect to the RIS. 

Capitalizing on the proposed near-field hierarchical codebook for the RIS, we devise an RIS profile and BS/MU beamforming alignment algorithm for establishing an RIS-enabled end-to-end multiple-input multiple-output channel according to a predefined performance objective.%, which is followed by pilot-assisted CE.  

\subsection{Near-Field Hierarchical Codebook Design}
\vspace{-0.1cm}
A convenient way to represent the RIS GRCS in the near-field case is to model the BS as a point source, which leads to the following expression for $g_{\rm ris}$ appearing in \eqref{eq:pathloss}:
	\begin{IEEEeqnarray}{ll}\label{Eq:gfunc_far}
		g_{\rm ris}(\bp_i,\bp_r)  
		= g\sum_{n=0}^{Q-1}
		\e^{\frac{\jj 2\pi}{\lambda}
		\big[\|\bp_i-\bp_{n}\|+\|\bp_r-\bp_{n}\|\big]}
		\e^{\jj\omega_{n}},\quad
	\end{IEEEeqnarray}
where $\bp_i$, $\bp_r$, and $\bp_n,\,\,\forall n$, denote the position of the BS, observation point, and the $n$-th RIS element, respectively. In contrast to the far-field case where the BS, MU, and the channel scatterers may belong to a wide range of angles of arrival and departure \cite{jamali2021power}, in the near-field case, it is more reasonable to focus on only the BS-RIS and RIS-MU LOS paths, where the RIS aims to illuminate the MU in a closely placed blockage area $\bp_r\in\sP_r$. Therefore, our objective is to design a finite set of phase-shift configurations which together are able to illuminate the entire blockage area. For simplicity, we assume that the blockage area is a rectangular one with its center at the point $\bp_b$, length $R_\x$, width $R_\y$, and the same height for all points, i.e., $\sP_r=\{\bp:\bp=\bp_b+[\x,\y,0],\,\,\forall\x\in[-R_\x/2,R_\x/2], \y\in[-R_\y/2,R_\y/2]\}$.

The problem of wide near-field illumination was recently treated in \cite{jamali2021low}, but not for a hierarchical codebook design. Exploiting this result, we propose the RIS phase-shift profile:
%------
\begin{IEEEeqnarray}{ll}\label{Eq:NearFieldPhase}
	\omega_{n} =
	-\frac{2\pi}{\lambda} \big[&\|\sM(\bp_n)-\bp_n\| \nonumber \\
	&- \|\sM(\bp_n)-\bp_{\rm ris}\|+\|\bp_i-\bp_{n}\|\big], \quad \forall n, \quad
\end{IEEEeqnarray}
%------
where $\bp_{\rm ris}$ represents the center of the RIS and $\sM(\bp_n)$ denotes a mapping from $\bp_n,\,\,\forall n$, to the points on the blockage area $\bp_r\in\sP_r$, which is given as follows:
%------
\begin{IEEEeqnarray}{ll}\label{Eq:Mapping}
\sM(\bp_n) = \bp_r+\left[\frac{\Delta_\x}{L_\z}\z+\x_{w_\x}, \frac{\Delta_\y}{L_\y}\y+\y_{w_\y},0\right]
\end{IEEEeqnarray}
%------
with $t_{w_t}\triangleq\frac{w_t R_t}{W_t}-R_t/2$ and $\Delta_t \triangleq \frac{\alpha R_t}{W_t}$, where $w_t=0,1,\dots,W_t-1$, $t\in\{\x,\y\}$. In \eqref{Eq:Mapping}, the term $[\x_{w_\x},\y_{w_\y},0]$ determines the center of illumination by the codeword $(w_\x,w_\y)$, and the parameter $(\Delta_\x,\Delta_\y)$ controls the illumination's width, where $\alpha$ specifies the degree of overlapping illumination among neighboring codewords. Note that, while the width of the illuminated area along the $\y$-axis is mainly due to the phase shifts applied by the RIS along the $\y$-axis, the width of the illuminated area along the $\x$-axis is mainly due to the phase shifts applied by the RIS along the $\z$-axis. The discritization of $[\x_{w_\x},\y_{w_\y},0]$ and the value of $(\Delta_\x,\Delta_\y)$ are chosen such that the entire blockage area $\bp_r\in\sP_r$ is covered via $W_\x W_\y$ RIS phase-shift configurations.

\begin{figure*}[t]
	%%%%%% Level 4
		\begin{minipage}{1\textwidth}
	\centering 
	 Level 4 ($|\sW_4|=256$)
\end{minipage}
		\begin{minipage}{0.49\textwidth}
	\centering 
	\includegraphics[width=1\columnwidth]{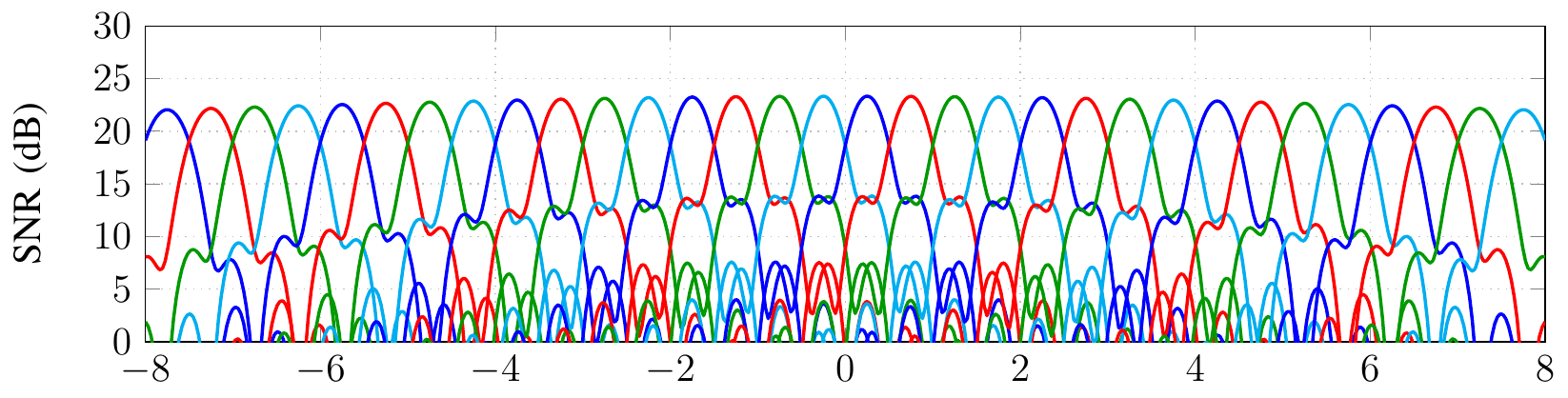}
\end{minipage}
		\begin{minipage}{0.49\textwidth}
	\centering 
	\includegraphics[width=1\columnwidth]{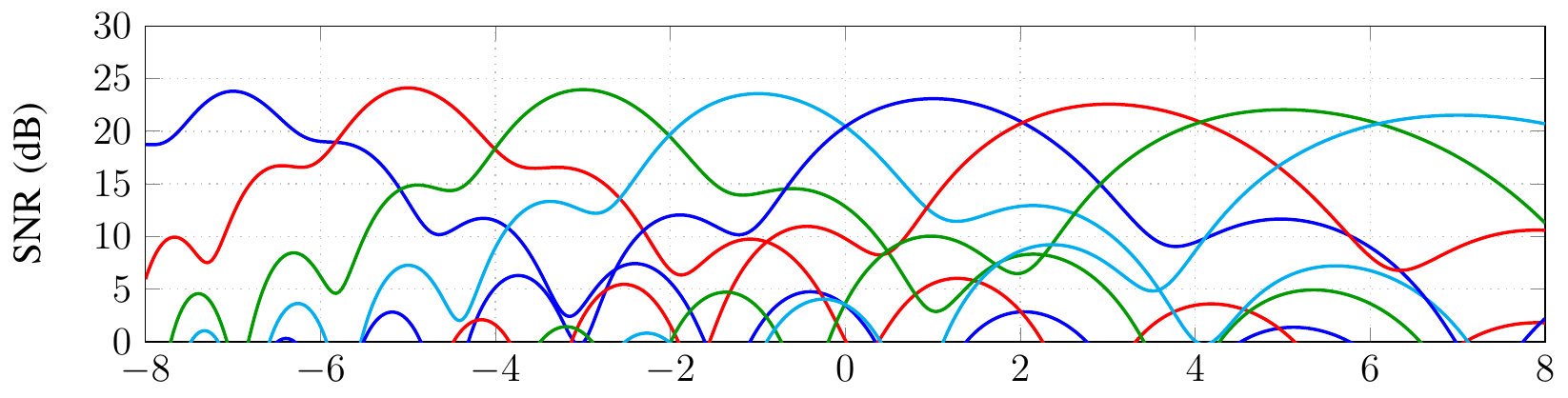}
		\end{minipage}
		%%%%%% Level 3
				\begin{minipage}{1\textwidth}
	\centering 
	 Level 3 ($|\sW_3|=128$)
\end{minipage}
		\begin{minipage}{0.49\textwidth}
	\centering 
	\includegraphics[width=1\columnwidth]{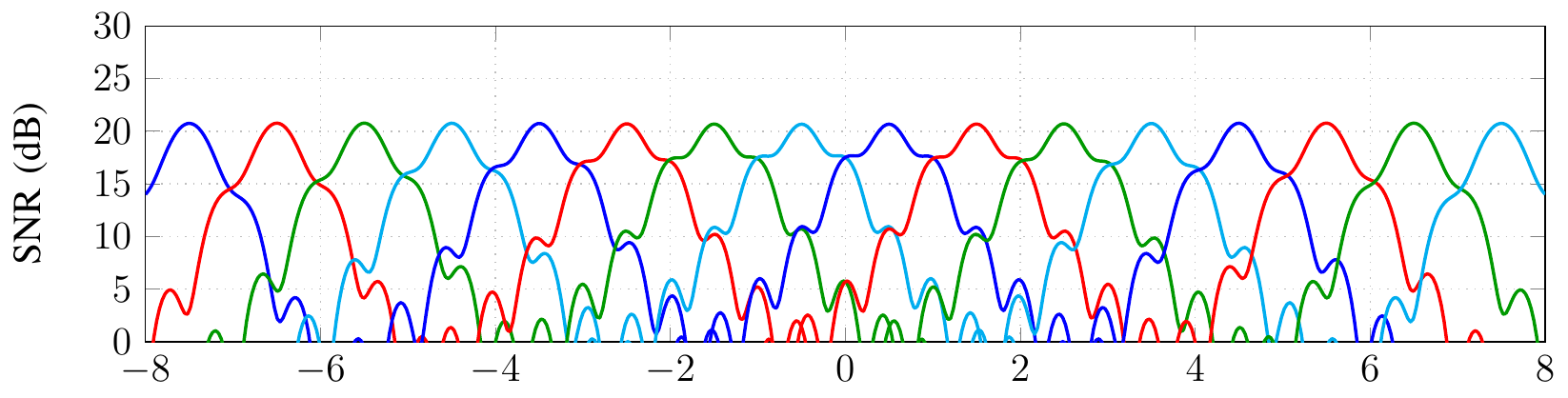}
\end{minipage}
		\begin{minipage}{0.49\textwidth}
	\centering 
	\includegraphics[width=1\columnwidth]{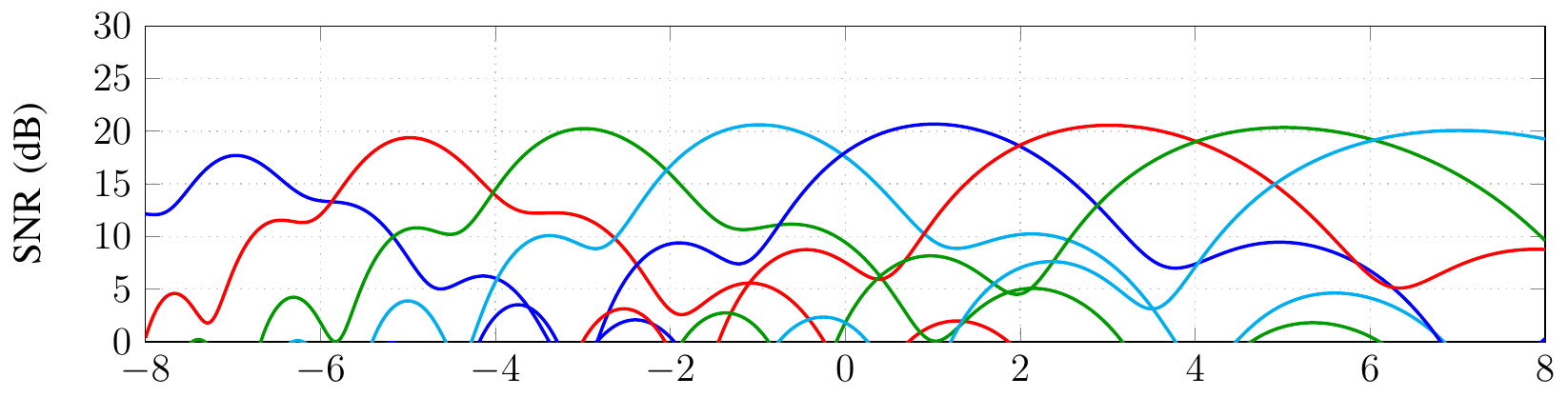}
		\end{minipage}
		%%%%%% Level 2
				\begin{minipage}{1\textwidth}
	\centering 
	 Level 2 ($|\sW_2|=64$)
\end{minipage}
		\begin{minipage}{0.49\textwidth}
	\centering 
	\includegraphics[width=1\columnwidth]{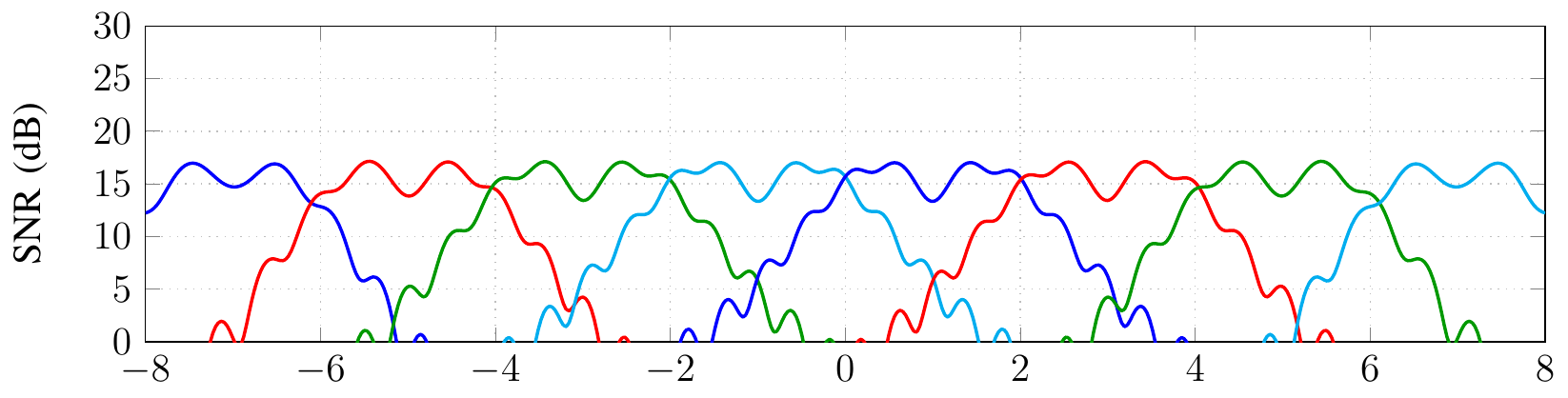}
\end{minipage}
		\begin{minipage}{0.49\textwidth}
	\centering 
	\includegraphics[width=1\columnwidth]{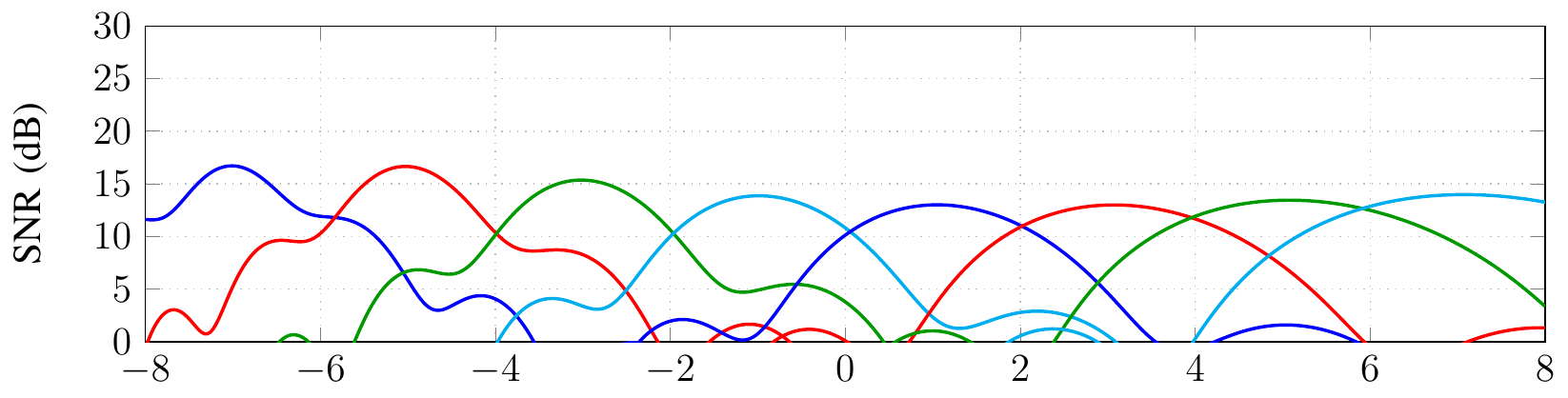}
		\end{minipage}
		%%%%%% Level 1
				\begin{minipage}{1\textwidth}
	\centering 
	 Level 3 ($|\sW_1|=16$)
\end{minipage}
		\begin{minipage}{0.49\textwidth}
	\centering 
	\includegraphics[width=1\columnwidth]{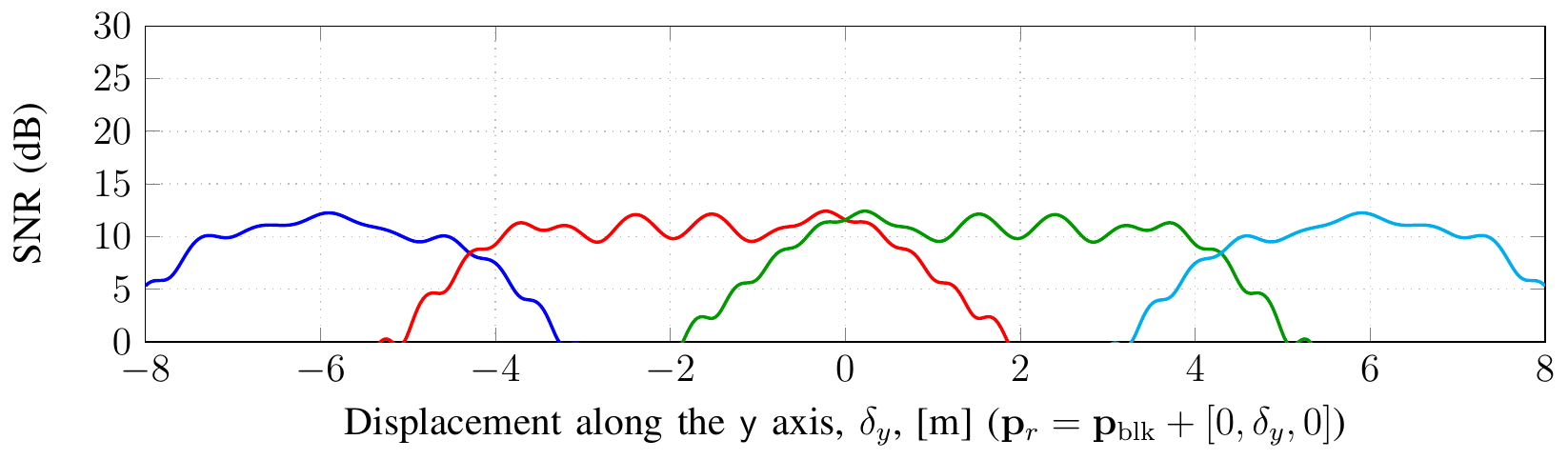}
\end{minipage}
		\begin{minipage}{0.49\textwidth}
	\centering 
	\includegraphics[width=1\columnwidth]{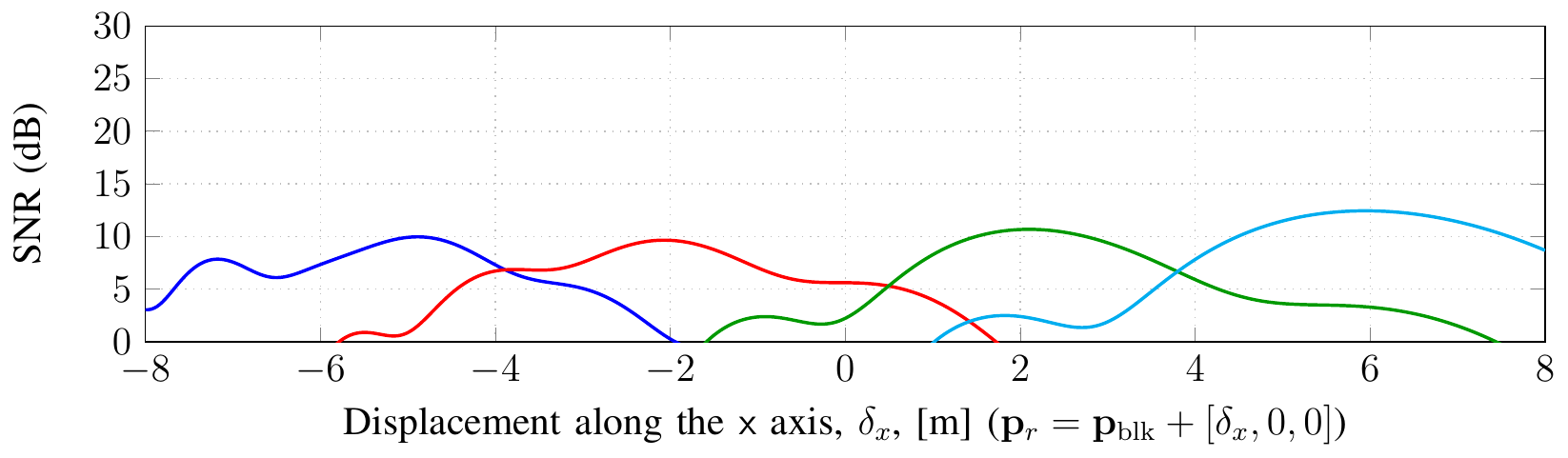}
		\end{minipage}
	\caption{Four-level hierarchical RIS phase-shift codebook for near-field illumination constructed from \eqref{Eq:NearFieldPhase} with $\alpha=0.8$. The $\x$- and $\y$-axes of the blockage area are respectively divided into $W_\x=4$ and $W_\y=4$ sub-areas for Level~1, $W_\x=8$ and $W_\y=8$ sub-areas for Level~8, $W_\x=8$ and $W_\y=16$ sub-areas for Level~3, and $W_\x=8$ and $W_\y=32$ sub-areas for Level~4. The system parameters are the same as those assumed in Section~IV and  Fig.~\ref{fig:focus}.}\vspace{-0.5cm}\label{fig:codebook_near}
\end{figure*}

It is crucial to notice that the values of ranges $W_\x$ and $W_\y$ are not necessarily identical due to the different illumination widths that they can generate along the $\x$ and $\y$-axes, respectively. To show this feature, we assume that the RIS focuses on point $\bp_{b}$ via the following focusing phase-shift profile:
%------
\begin{IEEEeqnarray}{ll}\label{Eq:Focus}
	\omega_{n} =\sum_{t\in\{\y,\z\}} 
	-\frac{2\pi d_t}{\lambda} \big[\|\bp_i-\bp_{n}\|+\|\bp_b-\bp_{n}\|\big],
	\quad \forall n.\quad
\end{IEEEeqnarray}
%------
This leads to the maximum GRCS $g_{\max}\triangleq\underset{\omega_{n}}{\max}|g_{\rm ris}(\bp_i,\bp_b)|=gQ$ at point $\bp_b$. In Fig.~\ref{fig:focus}, we
plot the Signal-to-Noise Ratio (SNR), defined as $\text{SNR}\triangleq\frac{P_{\rm bs}|g_{\rm ris}(\bp_i,\bp_r)|^2}{\sigma_{\rm mu}^2}$, of the RIS-enabled link as a function of the displacement  from $\bp_b$ along the $\x$-axis (denoted by $\delta_\x$) and the $\y$-axis (denoted by $\delta_\y$) (i.e., $\bp_r=\bp_b+[\delta_\x,0,0]$ and $\bp_r=\bp_b+[0,\delta_\y,0]$, respectively). As shown in the figure, the power distributions along the $\x$- and $\y$-axes on the blockage area exhibit a completely different behavior. In particular, for a given $\Delta_\x=\Delta_\y=\Delta$ and assuming that the blockage area is in front of the RIS (i.e., $\y_b=\y_{\rm ris}$), it is observed that the spread of the beams across the $\y$-axis is relatively regular. However, this is not the case for the $\x$-axis, where the beams become wider as we move away from the RIS. This is due to the fact that the wave propagates along the $\x$-axis, which further increases the spread of the beamwidth. This implies that for narrow illumination, $W_\y$ should be chosen larger than  $W_\x$ in order to cover the entire coverage area. Using this intuition, Fig.~\ref{fig:codebook_near} illustrates an example multi-level codebook design using \eqref{Eq:NearFieldPhase}, which includes ${\rm W}=4$ levels with respective sizes $(W_\x, W_\y)=(4,4),(8,8),(8,16),(8,32)$ for levels $w=1,2,3$, and $4$ (i.e., codebook sizes $16,64,128$, and $256$, respectively). We see from Fig.~\ref{fig:codebook_near} that for a large RIS with $Q=8649$, the proposed  codebook with size $W=256$ achieves a peak illumination SNR of $23$~dB at $\bp_s$ which is within only $2$~dB of the maximum SNR achieved via full focusing, see Fig.~\ref{fig:focus}.

%\section{Codebook-Based Channel Estimation}
\subsection{RIS Phase-Shift Management}
\vspace{-0.1cm}
%\subsection{Beam Management Strategy}
We consider simple BS precoding and MU combining designs and focus mainly on the design of RIS phase-shift profiles. Recall that the RIS is deployed in ways such that it has strong LOS connection to the BS and MU. Since the BS and RIS are fixed-position nodes, we consider a fixed precoder $\bv_{\rm bs}$ that is designed to focus on the RIS center. However, the MU may not know the position of the RIS, and hence, we assume that it can apply a set of $\bu_{\rm mu}$ combiners (denoted by $\sW_{\rm mu}$), choosing the one maximizing the received signal power.  By denoting as $\mathbf{\Omega}_{w,k_w}=g{\rm diag}\left(\e^{\jj\bomega_{w,k_w}}\right)$  the RIS phase-shift profile using the $k_w$-th codeword in the $w$-th level codebook $\sW_w$, the instantaneous SNR at the MU is calculated as follows:
\begin{IEEEeqnarray}{ll}\label{Eq:DL_Received_Signal_SNR}
\textsc{SNR}_{\rm mu}\!\left(\mathbf{\Omega}_{w,k_w}\right)=\!\!\!\! \underset{\mathbf{u}_{\rm mu}\in\sW_{\rm mu}}{\max}\!\!\!\!\frac{\left|\mathbf{u}_{\rm mu}^{\rm H}(\mathbf{H}+\mathbf{H}_2\mathbf{\Omega}_{w,k_w}\mathbf{H}_1)\mathbf{v}_{\rm bs}\right|^2}{\sigma_{\rm mu}^2},\quad
\end{IEEEeqnarray}
where we assumed $\|\mathbf{u}_{\rm mu}\|=1,\,\,\forall \mathbf{u}_{\rm mu}\in\sW_{\rm mu}$.  
During each SNR measurement, the BS sends pilot symbols, the RIS sets one phase-shift configuration $\mathbf{\Omega}_{w,k_w}$, and the MU empirically computes $\textsc{SNR}_{\rm mu}\left(\mathbf{\Omega}_{w,k_w}\right)$. Then, for each level of the RIS codebook, the MU reports the index of the phase shift with the maximum SNR, denoted by $k_w^*$, back to the RIS controller using a feedback control link. This is mathematically expressed as:
\begin{IEEEeqnarray}{ll}\label{Eq:MaxSNR}
k_w^*=\underset{k_w\in\mathcal{K}_w^*}{\rm argmax}\,\,\textsc{SNR}_{\rm mu}\left(\mathbf{\Omega}_{w,k_w}\right),
\end{IEEEeqnarray}
where $\mathcal{K}_w^*\subset\{1,2,\dots,|\mathcal{W}_w|\}$ is the reduced set of phase-shift indices used at the $w$-th codebook level. In particular, the RIS uses $k_w^*$ to determine $\mathcal{K}_{w+1}^*$ to consequently narrow down the search in the next codebook level. Obviously, the codebook at the first level is not reduced, i.e., $\mathcal{K}_1^*\triangleq\{1,2,\dots,|\mathcal{W}_1|\}$. The hierarchical RIS beam management is summarized in Algorithm~\ref{Alg:Beam}.

%%%%%%%%%%%%%%%%%%%%%%%%%%%%%%%%%%%%%%%%
%%            Algorithm 1           %%%
%%%%%%%%%%%%%%%%%%%%%%%%%%%%%%%%%%%%%%%%
\begin{algorithm}[t!]\caption{Hierarchical RIS Beam Management}\label{Alg:Beam}
\begin{algorithmic}[1]\small
\State \textbf{Input:} $\bu_{\rm bs}$, $\sW_{\rm mu}$, and $\sW_{\rm ris}=\sW_1\cup\sW_{2}\cup\cdots\cup\sW_{\rm W}$.
\For {$w=1,2\ldots,\mathrm{W}$}
\If{$w=1$}
\State Set $\mathcal{K}_1^*=\{1,2,\dots,|\mathcal{W}_1|\}$.
\Else
\State Determine $\mathcal{K}_w^*$ using $k^*_{w-1}$.
\EndIf
\For{$k_w\in\mathcal{K}_w^*$}
\State The RIS sets the phase shift $\mathbf{\Omega}_{w,k_w}$. 
\State The BS sends pilot symbols using precoder $\bv_{\rm bs}$.
\State The MU empirically computes the received SNR from \eqref{Eq:DL_Received_Signal_SNR} 

\hspace{0.27cm}using a combiner $\bu_{\rm mu}\in\mathcal{W}_{\rm mu}$.
\EndFor
\State The MU determines the best RIS phase shift from \eqref{Eq:MaxSNR} and 

\hspace{-0.2cm}reports its index $k_w^*$ back to the RIS controller.
\EndFor{}
\State \textbf{Output:} Optimized RIS phase-shift configuration $\bomega_{{\rm W},k_{\rm W}^*}$.
\end{algorithmic}
\end{algorithm}
%%%%%%%%%%%%%%%%%%%%%%%%%%%%%%%%%%%%%%%%

\vspace{-0.1cm}
\section{Numerical Results and Discussion}
\vspace{-0.1cm}
The simulation setup uses the following parameters. A carrier frequency $f=28$~GHz and half-wavelength antenna element spacing at all nodes were adopted. In particular, the BS is equipped with a square array of $N_{\rm bs}=64$ antennas placed on the $\x-\z$ plane, with the center at $\bp_{i} =[40, 0, 10]$. The RIS has a square aperture of size $L_\y=L_\z=0.5$~m (i.e., $Q=8649$) placed on the $\y-\z$ plane, with the center at $\bp_{\rm ris} =[0, 40, 5]$. Although the proposed algorithm is valid for multi-antenna MUs, in the simulation results, we assumed $N_{\rm mu}=1$ in order to compare our algorithm with Benchmark 3 (full CSI), which is introduced later. The MU is located in a square blockage area of size $\Delta_\x^{\max}\times\Delta_\y^{\max}$ parallel to the $\x-\y$ plane with center $\bp_{s} =[20, 40, 1]$, and we set $\Delta_\x^{\max}=\Delta_\y^{\max}=16$~m. The blockage in the direct BS-MU link is modeled by an attenuation factor of $20$~dB. Moreover, in addition to LOS paths, we assume that there are $20$ non-LOS scattering paths in the BS-RIS, BS-MU, and RIS-MU links, i.e., $\ell_{1}=\ell_{2}=\ell_{h}=21$, respectively. The positions of the channel scatters are randomly generated within volume $\mathcal{V}_s=\{(\x,\y,\z)|\x\in(0,60~\text{m}),\y\in(0,60~\text{m}),\z\in(0,10~\text{m})\}$. We assume $P_{\rm bs}=20$~dBm, noise spectral density $-176$~dB/Hz, bandwidth $B=100$~MHz, and  noise figure $N_0=6$~dB. The simulation results are generated for $100$ random realizations of the positions of the MU and the channel scatterers.  

We use the four-level hierarchical RIS phase-shift codebook presented in Fig.~\ref{fig:codebook_near}. Therefore, the BS has to send $16$, $4$, $2$, and $2$ pilots at codebook levels $1$, $2$, $3$, and $4$, respectively. The proposed algorithm relies on the following three design choices: \textit{i)} hierarchical search over \textit{ii)} phase-shift configuration codebooks \textit{iii)} designed based on LoS paths. Therefore, we choose the following benchmarks to evaluate the efficiency of these options. \textbf{Benchmark~1 (full codebook search):} Here, we consider the largest codebook (i.e., Level $4$ with $256$ phase-shift configurations) and perform a full search to find the best phase shift. \textbf{Benchmark~2 (full focusing):} We assume that the MU position is perfectly known and the RIS designs its phase shifts to focus on it, cf. \eqref{Eq:Focus}. \textbf{Benchmark~3 (full CSI):} For a fixed BS precoder and single-antenna MU, the effective RIS-enabled channel reduces to $\mathbf{h}_2\mathbf{\Omega}\mathbf{h}_1$, where $\mathbf{h}_1\triangleq\mathbf{H}_1\bv_{\rm bs}$ and $\mathbf{h}_2$ is the RIS-MU channel vector. In this case, assuming full CSI (i.e., $\mathbf{h}_1$ and $\mathbf{h}_2$) is available, the optimal phase shifts to maximize the end-to-end channel gain are $[\mathbf{\Omega}]_q=-[\angle \mathbf{h}_1]_q-[\angle \mathbf{h}_2]_q,\,\,\forall q$. Benchmarks~1-3 constitute performance upper bounds for the proposed RIS beam management algorithm and can be used to access the three considered design choices.

\begin{figure*}[t]
		\centering
		\includegraphics[width=1.4\columnwidth]{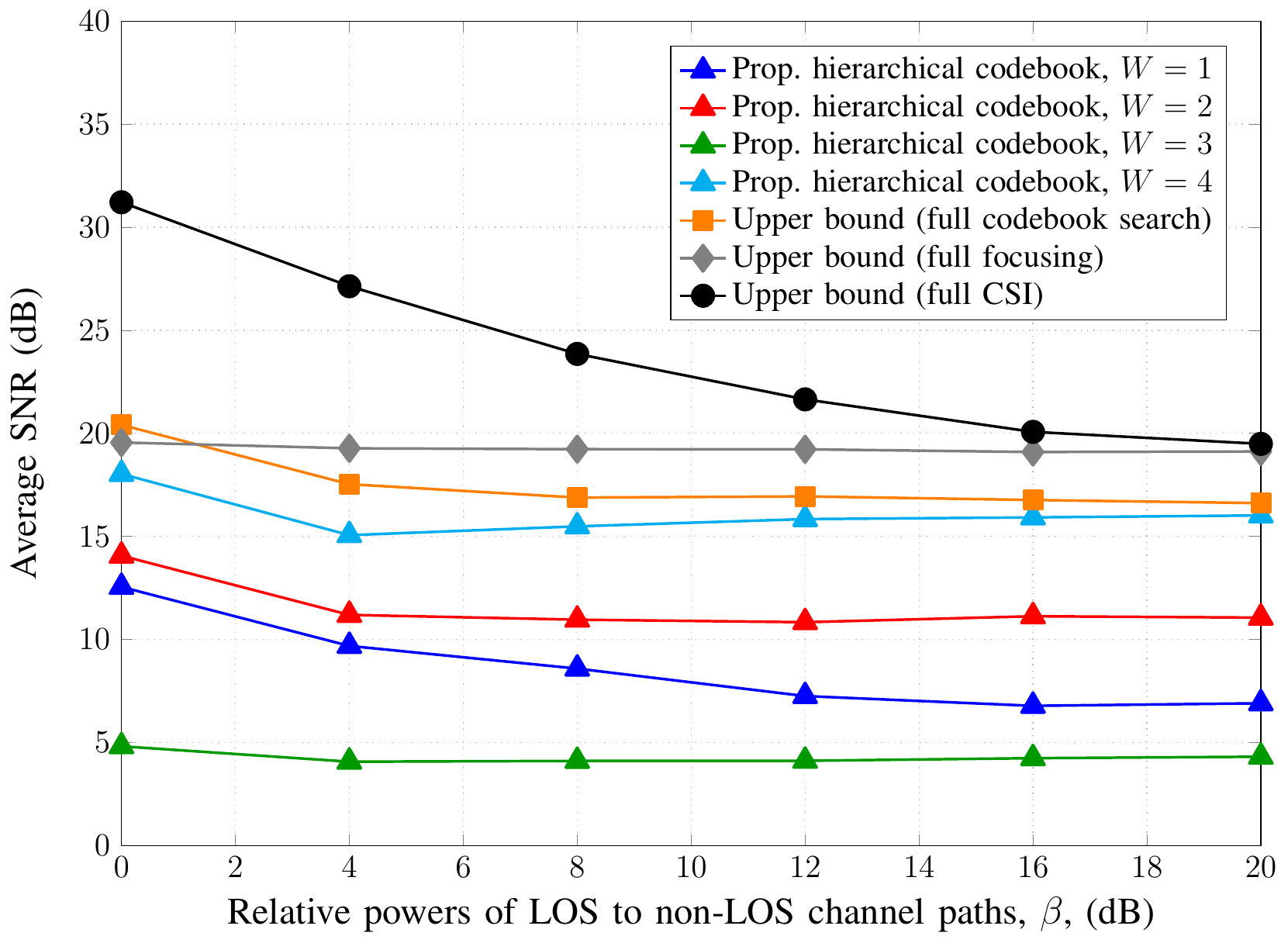}\hspace{-0.18cm}
		\caption{Average received SNR at the MU (dB) vs. the relative powers of the LOS path with respect to the non-LOS paths (dB).}\vspace{-0.5cm}\label{fig:rate}	
\end{figure*}

In Fig.~\ref{fig:rate}, we plot the average received SNR (dB) vs. the relative powers (dB) of the LOS and non-LOS paths, defined as $\beta=\frac{\text{PL}_{t,0}}{\sum_{i=1}^{\ell_t-1}\text{PL}_{t,i}}$ with $t\in\{h,1,2\}$. We assume free-space path loss for the LOS paths and change the relative power of non-LOS paths to vary $\beta$. As expected, for almost all values of $\beta$, the schemes can be sorted as Benchmark~3, Benchmark~2, Benchmark~1, and the proposed algorithm in descending order according to their performance. In particular, Benchmark~3 considerably outperforms the other schemes if the non-LOS paths are dominant (which is an unlikely scenario for mmWave communication systems \cite{ULBA2021}). Whereas, at LOS-dominated scenarios, the performance of Benchmark~3 approaches that of the remaining schemes, which are designed based on the LOS path. In fact, except Benchmark~3, the performance of the other schemes is relatively stable across all the considered range of $\beta$ values. The reason for this behavior is that, in the mmWave bands, the transmission beams are very narrow, which suggests that despite the large number of channel scatterers, very few may fall into the transmission beams. Assuming $\beta=10$~dB, which is practical for mmWave systems, Benchmark~3, Benchmark~2, Benchmark~1, and the proposed algorithm achieve average SNRs of $25$, $19$, $17$, and $16$ dB, respectively. However, to achieve these performances, Benchmark~3 needs $\mathbf{h}_1$ and $\mathbf{h}_2$ containing an overall of $2\times Q=17298$ channel coefficients; Benchmark~2 requires perfect MU position knowledge, requiring very high-resolution localization; Benchmark~1 needs $256$ pilot transmissions; whereas the proposed algorithm requires only $24$ pilot transmissions. \textbf{The high performance achieved with such an extremely low overhead control, compared to the considered benchmark schemes, implies a high practical appeal of the proposed beam management algorithm for RIS-enabled mmWave communication systems.} 
%\section{Conclusion}

\scriptsize{
\section*{Acknowledgments}
The work of G. C. Alexandropoulos was supported by the EU H2020 RISE-6G project under grant number 101017011; the work of R. Schober was supported in part by the DFG project SCHO 831/15-1 and the BMBF project 6G-RIC with ID 16KISK023; and the work of H. V. Poor was supported by the U.S. National Science Foundation under grant CCF-1908308.}

\bibliographystyle{IEEEtran}
\bibliography{References}

\begin{thebibliography}{10}
\providecommand{\url}[1]{#1}
\csname url@samestyle\endcsname
\providecommand{\newblock}{\relax}
\providecommand{\bibinfo}[2]{#2}
\providecommand{\BIBentrySTDinterwordspacing}{\spaceskip=0pt\relax}
\providecommand{\BIBentryALTinterwordstretchfactor}{4}
\providecommand{\BIBentryALTinterwordspacing}{\spaceskip=\fontdimen2\font plus
\BIBentryALTinterwordstretchfactor\fontdimen3\font minus
  \fontdimen4\font\relax}
\providecommand{\BIBforeignlanguage}[2]{{%
\expandafter\ifx\csname l@#1\endcsname\relax
\typeout{** WARNING: IEEEtran.bst: No hyphenation pattern has been}%
\typeout{** loaded for the language `#1'. Using the pattern for}%
\typeout{** the default language instead.}%
\else
\language=\csname l@#1\endcsname
\fi
#2}}
\providecommand{\BIBdecl}{\relax}
\BIBdecl

\bibitem{liaskos2018new}
C.~Liaskos, S.~Nie, A.~Tsioliaridou, A.~Pitsillides, S.~Ioannidis, and I.~F.
  Akyildiz, ``A new wireless communication paradigm through software-controlled
  metasurfaces,'' \emph{IEEE Commun. Mag.}, vol.~56, no.~9, pp. 162--169, Sep.
  2018.

\bibitem{di2019smart}
M.~Di~Renzo, M.~Debbah, D.-T. Phan-Huy, A.~Zappone, M.-S. Alouini, C.~Yuen,
  V.~Sciancalepore, G.~C. Alexandropoulos, J.~Hoydis, H.~Gacanin, J.~de~Rosny,
  A.~Bounceu, G.~Lerosey, and M.~Fink, ``Smart radio environments empowered by
  {AI} reconfigurable meta-surfaces: {An} idea whose time has come,''
  \emph{EURASIP J. Wireless Commun. Netw.}, vol. 129, May 2019.

\bibitem{wu2019towards}
Q.~Wu and R.~Zhang, ``Towards smart and reconfigurable environment: Intelligent
  reflecting surface aided wireless network,'' \emph{IEEE Commun. Mag.},
  vol.~58, no.~1, pp. 106--112, Jan. 2020.

\bibitem{RISE6G_2021}
E.~Calvanese~Strinati, G.~C. Alexandropoulos, H.~Wymeersch, B.~Denis,
  V.~Sciancalepore, R.~D'Errico, A.~Clemente, D.-T. Phan-Huy, E.~D. Carvalho,
  and P.~Popovski, ``Reconfigurable, intelligent, and sustainable wireless
  environments for {6G} smart connectivity,'' \emph{IEEE Commun. Mag.},
  vol.~59, no.~10, pp. 99--105, Oct. 2021.

\bibitem{huang2019reconfigurable}
C.~Huang, A.~Zappone, G.~C. Alexandropoulos, M.~Debbah, and C.~Yuen,
  ``Reconfigurable intelligent surfaces for energy efficiency in wireless
  communication,'' \emph{IEEE Trans. Wireless Commun.}, vol.~18, no.~8, pp.
  4157--4170, Aug. 2019.

\bibitem{WavePropTCCN}
G.~C. Alexandropoulos, G.~Lerosey, M.~Debbah, and M.~Fink, ``Reconfigurable
  intelligent surfaces and metamaterials: {T}he potential of wave propagation
  control for {6G} wireless communications,'' \emph{IEEE ComSoc TCCN
  Newslett.}, vol.~6, no.~1, pp. 25--37, Jun. 2020.

\bibitem{risTUTORIAL2020}
Q.~Wu, S.~Zhang, B.~Zheng, C.~You, and R.~Zhang, ``Intelligent reflecting
  surface aided wireless communications: {A} tutorial,'' \emph{IEEE Trans.
  Commun.}, vol.~69, no.~5, pp. 3313--3351, May 2021.

\bibitem{wei2021channel}
L.~Wei, C.~Huang, G.~C. Alexandropoulos, C.~Yuen, Z.~Zhang, and M.~Debbah,
  ``Channel estimation for {RIS}-empowered multi-user {MISO} wireless
  communications,'' \emph{IEEE Trans. Commun.}, vol.~69, no.~6, pp. 4144--4157,
  Jun. 2021.

\bibitem{yu2021smart}
X.~Yu, V.~Jamali, D.~Xu, D.~W.~K. Ng, and R.~Schober, ``Smart and
  reconfigurable wireless communications: {From} {IRS} modeling to algorithm
  design,'' \emph{IEEE Wireless Commun.}, vol.~28, no.~6, pp. 118--125, Dec.
  2021.

\bibitem{CE_overview_2022}
M.~Jian, G.~C. Alexandropoulos, E.~Basar, C.~Huang, R.~Liu, Y.~Liu, and
  C.~Yuen, ``Reconfigurable intelligent surfaces for wireless communications:
  {O}verview of hardware designs, channel models, and estimation techniques,''
  \emph{ITU Intell. Converged Netw.}, to appear, 2022, [Online]
  https://arxiv.org/abs/2203.03176.

\bibitem{wang2020compressed}
P.~Wang, J.~Fang, H.~Duan, and H.~Li, ``Compressed channel estimation for
  intelligent reflecting surface-assisted millimeter wave systems,'' \emph{IEEE
  Sig. Process. Lett.}, vol.~27, pp. 905--909, 2020.

\bibitem{jamali2021power}
V.~Jamali, M.~Najafi, R.~Schober, and H.~V. Poor, ``Power efficiency, overhead,
  and complexity tradeoff of {IRS} codebook design--{Quadratic} phase-shift
  profile,'' \emph{{IEEE} Commun. Lett.}, vol.~25, no.~6, pp. 2048--2052, Jun.
  2021.

\bibitem{Fast_Beam_Rui_2020}
C.~You, B.~Zheng, and R.~Zhang, ``Fast beam training for {IRS}-assisted
  multiuser communications,'' \emph{IEEE Wireless Commun. Lett.}, vol.~9,
  no.~11, pp. 1845--1849, Nov. 2020.

\bibitem{Hierarchical_mmWave_2016}
Z.~Xiao, T.~He, P.~Xia, and X.-G. Xia, ``Hierarchical codebook design for
  beamforming training in millimeter-wave communication,'' \emph{IEEE Trans.
  Wireless Commun.}, vol.~15, no.~5, pp. 3380--3392, May 2016.

\bibitem{Singh_2021}
C.~Singh, K.~Singh, and K.~H. Liu, ``Fast beam training for {RIS}-assisted
  uplink communication,'' Jul. 2021, [Online] https://arxiv.org/pdf/2107.14138.

\bibitem{Ning_THz_2021}
B.~Ning, Z.~Chen, W.~Chen, Y.~Du, and J.~Fang, ``Terahertz multi-user massive
  {MIMO} with intelligent reflecting surface: {B}eam training and hybrid
  beamforming,'' \emph{IEEE Trans. Veh. Technol.}, vol.~70, no.~2, pp.
  1376--1393, Feb. 2021.

\bibitem{Wang_2021}
W.~Wang and W.~Zhang, ``Joint beam training and positioning for intelligent
  reflecting surfaces assisted millimeter wave communications,'' \emph{IEEE
  Trans. Wireless Commun.}, vol.~20, no.~10, pp. 6282--6297, Oct. 2021.

\bibitem{Hu_multicast_2020}
X.~Hu, C.~Zhong, Y.~Zhu, X.~Chen, and Z.~Zhang, ``Programmable
  metasurface-based multicast systems: {D}esign and analysis,'' \emph{IEEE J.
  Sel. Areas Commun.}, vol.~38, no.~8, pp. 1763--1776, Aug. 2020.

\bibitem{Wang_THz_2021}
P.~Wang, J.~Fang, W.~Zhang, and H.~Li, ``Fast beam training and alignment for
  {IRS}-assisted millimeter wave/terahertz systems,'' Oct. 2021, [Online]
  https://arxiv.org/abs/2103.05812.

\bibitem{cai_Hierarchical_2021}
C.~Cai, X.~Yuan, W.~Yan, Z.~Huang, Y.-C. Liang, and W.~Zhang, ``Hierarchical
  passive beamforming for reconfigurable intelligent surface aided
  communications,'' \emph{IEEE Wireless Commun. Lett.}, vol.~10, no.~9, pp.
  1909--1913, Sep. 2021.

\bibitem{zegrar_General_2020}
S.~E. Zegrar, L.~Afeef, and H.~Arslan, ``A general framework for {RIS}-aided
  {mmWave} communication networks: {Channel} estimation and mobile user
  tracking,'' Sep. 2020, [Online] https://arxiv.org/abs/2009.01180.

\bibitem{jamali2021low}
V.~Jamali, G.~C. Alexandropoulos, R.~Schober, and H.~V. Poor,
  ``Low-to-zero-overhead {IRS} reconfiguration: {Decoupling} illumination and
  channel estimation,'' \emph{IEEE Commun. Lett.}, early access, 2022.

\bibitem{balanis2015antenna}
C.~A. Balanis, \emph{Antenna theory: Analysis and design}.\hskip 1em plus 0.5em
  minus 0.4em\relax John wiley \& sons, 2015.

\bibitem{najafi2020intelligent}
M.~Najafi, V.~Jamali, R.~Schober, and H.~V. Poor, ``Physics-based modeling and
  scalable optimization of large intelligent reflecting surfaces,'' \emph{IEEE
  Trans. Commun.}, vol.~69, no.~4, pp. 2673--2691, Apr. 2021.

\bibitem{ULBA2021}
G.~C. Alexandropoulos, I.~Vinieratou, M.~Rebato, L.~Rose, and M.~Zorzi,
  ``Uplink beam management for millimeter wave cellular {MIMO} systems with
  hybrid beamforming,'' in \emph{Proc. {IEEE WCNC}}, Nanjing, China, Apr. 2021.

\end{thebibliography}

\end{document}